\documentclass[12pt]{article}

\def\fnote#1#2{\begingroup\def\thefootnote{#1}\footnote{#2}\addtocounter{footnote}{-1}\endgroup}

\usepackage{amssymb}
\usepackage{makeidx}
\usepackage{epsf}
\usepackage{graphicx}        

\makeindex

\def\fnote#1#2{\begingroup\def\thefootnote{#1}\footnote{#2}\addtocounter{footnote}{-1}\endgroup}

\def\inbar{\vrule height1.5ex width.4pt depth0pt}
\def\IB{\relax{\rm I\kern-.18em B}}
\def\IC{\relax\,\hbox{$\inbar\kern-.3em{\rm C}$}}
\def\ID{\relax{\rm I\kern-.18em D}}
\def\IE{\relax{\rm I\kern-.18em E}}
\def\IF{\relax{\rm I\kern-.18em F}}
\def\IG{\relax\,\hbox{$\inbar\kern-.3em{\rm G}$}}
\def\IH{\relax{\rm I\kern-.18em H}}
\def\II{\relax{\rm I\kern-.18em I}}
\def\IK{\relax{\rm I\kern-.18em K}}
\def\IL{\relax{\rm I\kern-.18em L}}
\def\IM{\relax{\rm I\kern-.18em M}}
\def\IN{\relax{\rm I\kern-.18em N}}
\def\IO{\relax\,\hbox{$\inbar\kern-.3em{\rm O}$}}
\def\IP{\relax{\rm I\kern-.18em P}}
\def\IQ{\relax\,\hbox{$\inbar\kern-.3em{\rm Q}$}}
\def\IR{\relax{\rm I\kern-.18em R}}
\def\IT{\relax{\rm I\kern-.18em T}}
\def\ZZ{\relax{\sf Z\kern-.4em Z}}

\def\a{\alpha}       \def\g{\gamma}  
\def\e{\epsilon} \def\G{\Gamma}     
     \def\si{\sigma}

\def\cA{{\cal A}}  \def\cC{{\cal C}}
  
 \def\cH{{\cal H}} \def\cI{{\cal I}}

\def\cP{{\cal P}}

     \def\veck{{\vec{k}}}

          \def\rmIm{{\rm Im}}

\def\rmNL{{\rm NL}}

 \def\rmPl{{\rm Pl}}

            \def\rmSL{{\rm SL}}
 \def\rmSO{{\rm SO}}

       \def\rmad{{\rm ad}}   
                \def\rmaut{{\rm aut}}

     \def\rmdet{{\rm det}}

\def\rmth{{\rm th}}            
 \def\rmtr{{\rm tr}}

\def\afrak{{\mathfrak a}} 
 \def\gfrak{{\mathfrak g}}   \def\kfrak{{\mathfrak k}}
 
\def\nfrak{{\mathfrak n}}   \def\pfrak{{\mathfrak p}}

 \def\mathR{{\mathbb R}}
\def\mathZ{{\mathbb Z}}

\def\fnote#1#2{\begingroup\def\thefootnote{#1}\footnote{#2}\addtocounter{footnote}{-1}\endgroup}

\def\beq{\begin{equation}}
\def\eeq{\end{equation}}
\def\bea{\begin{eqnarray}}
\def\eea{\end{eqnarray}}

\def\notin{\ \hbox{{$\in$}\kern-.51em\hbox{/}}}

\def\del{\partial}

  \def\E1Fq{E_1/\IF_q}

\def\notdiv{{\relax{~|\kern-.34em /~}}}

\def\of{{\overline{f}}}
\def\oj{{\overline{j}}}

   \def\ophi{{\overline{\phi}}}

\def\otau{{\overline{\tau}}}

\begin{document}

\phantom{\hfill \today}

\vskip 1truein
\parskip=0.15truein
\baselineskip=19pt

\centerline{\Large {\bf Automorphic Inflation }}

\vskip .2truein

\centerline{\sc Rolf Schimmrigk\fnote{1}{
  netahu@yahoo.com; rschimmr@iusb.edu}}

\vskip .2truein

\centerline{Dept. of Physics}

\centerline{Indiana University South Bend}

\centerline{1700 Mishawaka Ave., South Bend, IN 46634}

\vskip 1truein

 \centerline{\bf Abstract}

\baselineskip=17pt

\begin{quote}
 A framework of inflation is formulated based on symmetry groups and their
 associated automorphic functions.
 In this setting the inflaton multiplet takes values in a curved target space constructed
 from a continuous group $G$ and a discrete subgroup $\G$. The dynamics of
 inflationary models is essentially determined by the choice of the pair $(G,\G)$ and
 a function $\Phi$ on the group $G$ relative to $\G$.
 Automorphic inflation provides a natural structure in which the shift symmetry of
 large field inflation arises as one of generators of $\G$.
 The model of $j-$inflation is discussed  as an example of modular
 inflation associated to the special linear group.

\end{quote}

\renewcommand\thepage{}
\newpage
\pagenumbering{arabic}

\baselineskip=18pt
\parskip=0.05truein

\tableofcontents
 \vskip .8truein

\parindent=0pt
\parskip=0.15truein
\baselineskip=22pt

\section{Introduction}

 Symmetries have been useful as guides to the dynamics of fundamental theories for more than
 a century. The most dramatic examples involve continuous groups, but discrete groups have been
 of importance as well. The purpose of this paper is to formulate a field theoretic framework
 based on the combination of both types of groups as well as an associated function space.
 Given a continuous Lie group $G$ and a discrete subgroup $\G$ of $G$, the dynamics
 is determined by the choice of automorphic forms $\Phi_i$, defined relative to the discrete group
 $\G$ as a function on $G$.
 These forms descend to functions $f_i$ on the target space $X$ of the inflaton
 multiplet $\phi^I$, thereby inducing a potential $V(f_i(\phi^I))$.  The kinetic term
  of the resulting theory leads to a non-flat
  metric $G_{IJ}$, determined by the Lie algebra  of the group $G$.

 In this letter the framework of automorphic field theory is applied to inflation. This is
 motivated by issues that arise in models in which inflation occurs at a high
 energy scale.
 Estimates of the effects of higher dimension operators expected to appear
 in the UV-completion of inflationary models suggest that such
 operators can make model-specific predictions unstable.
 This is particularly pronounced in the framework of large field inflation, where
 the inflaton varies over an energy range that is super-Planckian.
 Unless these operators have very small coefficients, higher dimension corrections
 will have dramatic effects  on the parameters.
 An early discussion of these issues in the context of chaotic inflation
 \cite{l83} can be found in ref. \cite{em86}.
 A device often postulated to avoid such corrections
 is the existence of an inflaton shift symmetry $\phi \longmapsto \phi + s$.
  Historically, the first model to incorporate this idea is natural inflation
 \cite{freese-etal}, but many modifications and extensions have been introduced in the
 intervening two decades, including models that aim at realizations of this symmetry in
 UV-complete theories \cite{s08etal, pp13rev, s13rev}. This idea has received renewed
 attention following
 the possibility  of an observable tensor ratio  \cite{bicep2-14, fhs14, ms14}.
 If a sizable portion of  the CMB power spectrum comes from primordial gravitational waves
 the inflationary scale is quite high, not too far from that of GUT
 models \cite{l97}.

In the context of the shift symmetry it is natural to ask whether it is part of a larger group that
operates on the inflaton field space. The existence of such a group would provide a  systematic
 framework in which different  types of invariant potentials could be classified.
In the present letter such a program is initiated by formulating inflation in terms
 of symmetry groups and their associated automorphic functions. There is a historical
 precedent for such a strategy in quantum field theory, where the interpretation of
 an inversion symmetry as a generator of an infinite group led to the concept of
 duality.
 In the present framework the shift symmetry becomes part of an infinite
 subgroup $\G$ of $G$ with constraints strong enough to characterize
the space of forms.

\section{Automorphic inflation framework}

The structure of automorphic inflation is characterized in essence
by a potential determined in terms of automorphic forms, and a
metric $G_{IJ}$ of the kinetic term that is determined by the
symmetry group.

\subsection{Automorphic actions}

In the simplest case the concept of automorphic inflation can be formulated in the context
  of multi-field theories defined by an action of the form
 \beq
  \cA_\rmaut = \int d^4x \sqrt{-g} \left(\frac{M_\rmPl^2}{2}R
      - \frac{1}{2}G_{IJ}  \del_\mu \phi^I \del^\mu \phi^J
    - V(\phi^I)\right),
 \eeq
 where $M_\rmPl = 1/\sqrt{8\pi G}$ is the reduced Planck mass
 and the spacetime metric $g_{\mu \nu}$ is taken to be of signature $(-,+,+,+)$.
 The basic building blocks include potentials $V(\phi^I)$ induced by
 automorphic forms  $f$ on the inflaton field space
  $X$ with coordinates $\tau = (\tau^1,...,\tau^n)$ as
 $V(f(\tau^I)) = \Lambda^4 F(f(\tau^I))$.
Here $\Lambda$ is an energy scale, and  $F(\tau^I)$ is a
dimensionless function of the inflaton multiplet expressed in
terms of dimensionless variables $\tau = \phi/\mu$,
 where $\mu$ is a second energy scale. The potential function $F$ should be a real
 function $F(f,\of)$, a  simple class of examples given by $F(f,\of) = (f \of)^p$
  for arbitrary exponents $p$. Inflationary
 models $\cI$ of this type are thus characterized by a number of choices that characterize
the groups $(G,\G)$, the functions $(\Phi, F)$, and the energy
scales $(\Lambda,\mu)$. The space $X$ is to be taken as a bounded
region in the complex vector space of dimension $n$, whose
structure is constrained by the nature of the inflaton $\phi^I$.
 These spaces have a curved metric $G_{IJ}, I,J=1,...,n$,
 which will be described in more detail further below, after the nature of the space $X$ has
been made more explicit.

Automorphic inflation  can be formulated in a more general context
for
 theories of the form $P(E,V)$, where
  $E=G_{IJ} g^{\mu \nu} \del_\mu \phi^I \del_\nu \ophi^J$, and $P$ is some general,
 not necessarily polynomial, function. Examples of this type include the
 extension of DBI inflation \cite{dbi-inflation, l10}  to the multi-field context.

\subsection{Automorphic forms associated to groups}

The concept of automorphic forms was introduced in the second half
of the 19$^\rmth$ century in the context of functions on the
complex plane. This framework is too narrow
 for multi-field inflation, but the generalization of the
 classical notion of automorphic forms has made the
 concept less precise and no standard language has emerged.
In the following discussion automorphic forms are understood to be defined on
 higher rank matrix groups, to be distinguished from modular forms.
 Roughly speaking, forms of this type are highly symmetric functions on a continuous group $G$
 that are  characterized by their transformation behavior with respect to certain subgroups,
 and by  the fact that they are eigenfunctions of differential operators.
 In this way they can be viewed as objects reminiscent of tensors, but more rigid.
When $G$ is a semisimple Lie group with finite center
 there are essentially only two subgroups to consider, the discrete subgroup
 $\G$ and the maximal compact subgroup $K$. The
 inflaton space $X$ is defined as the quotient space $X=G/K$, which
 inherits a Riemannian metric from $G$. It is therefore natural to consider eigenfunctions
 of those differential operators that are invariant under the group action.

In more detail, the transformation behavior of the function
   $f$ on $X$ with respect to the discrete group $\G$
is determined by an automorphy factor $J(g, x)$ that depends both
on group elements $g$ and on elements $x$ of $X$.
 This function $J$ is defined by the relation $J(gh,x) = J(g,hx)J(h,x)$
 for $g,h\in G$ and $x\in X$. A $J$-automorphic form $f(x)$ is induced by
 a certain type of group function $\Phi(g)$ as
 \beq
 f(x) = J(g,x_0) \Phi(g),
\eeq
 where $x=g\cdot x_0$ and $x_0$ is a point left invariant by
 the subgroup $K$.
 In order to obtain the standard transformation behavior as
 \beq
 f(\g x)= \e(\g) J(\g, x) f(x),
\eeq
 where $\g$ is an element of $\G$ and $x$ is in $X$, the functions $\Phi$ have to satisfy a number of
constraints that are mostly concerned with the transformation behavior of $\Phi$ with respect
 to different kinds of subgroups of the group $G$. These constraints will be illustrated more
concretely  in the special case of modular forms further below.

 If $G$ is a semisimple Lie group with finite center
the first condition specifies the behavior of the function $\Phi$
with respect to the action
 of the discrete group $\G$ on $\Phi$, which in general is allowed to
 transform with a character $\e$ as  $\Phi(\g g) = \e(\g) \Phi(g)$, for $\g \in \G$.
The second constraint restricts the behavior of $\Phi$ with respect to the action of the
 maximal compact subgroup $K$ by requiring that the forms span a finite-dimensional
space under this action.
 The third condition, structurally different in type from the transformation constraints  above,
generalizes  the  eigenvalue constraints of classical modular
forms, either of holomorphic or non-holomorphic type. In the
higher rank case the function $\Phi$ is assumed
 to be an element of a finite dimensional eigenspace
  of all differential operators that are invariant with respect to the group
 structure. The number of generating operators of this type is finite, given by the rank of the
 group $G$.
 Finiteness results for automorphic forms have been established by imposing
  a convergence constraint that requires the existence of
   a constant $C$ and an integer $m$ such that
   $|\Phi(g)| \leq C ||g||^m$.  More details concerning the conceptual framework of
 automorphic forms
 can be found in the reviews of ref. \cite{bc79}.

The notion of automorphic inflation just outlined in terms of the group theoretic framework
  associates different models to each Lie group  $G$ and the subgroup $\G$
 via the choice of an automorphic form $\Phi$ and the function $F$.
 Once a symmetric bounded domain has been chosen,
 together with $\G$, the space of automorphic
 forms is finite-dimensional \cite{hc68, l76}. This implies that modulo the function $F$
 each choice $(G,  \G)$ leads to a  finite-dimensional theory space.

\subsection{The kinetic term of automorphic inflation}

The kinetic term of automorphic field theory
 is characterized by a nontrivial metric on the target space
 $X=G/K$ of the inflaton multiplet. This metric is induced in terms of the adjoint
 representation of the Lie algebra $\gfrak$, defined as
 $\rmad_V(W) = [V,W]$, via the Cartan-Killing form
 $B(X,Y) = \rmtr~\rmad_X \rmad_Y$ on  $\gfrak$ of $G$, which is isomorphic
 to the tangent space $T_eG$ at the identity element $e$.  The inner product
 on $T_eG$ can be transported
 to other tangent spaces $T_gG$ by the differential $dL_g$ of the left translation map $L_g$.
The pullback via $L_{g^{-1}}$ can be used to define the inner product on
 $T_gG$ as
 $\langle V,W\rangle_g = \langle dL_{g^{-1}}V, dL_{g^{-1}}W\rangle_e$,
 for tangent vectors $V,W$. The associated metric descends to the
 quotient $X=G/K$.

The coordinate form $G_{IJ}$ of the metric on $X$ can be obtained explicitly from
 the Iwawasa decomposition $G=NAK$ of $G$, defined as a refinement
  of the Cartan decomposition $\gfrak = \kfrak \oplus \pfrak$,
 where $\kfrak$ is the Lie algebra of the maximal compact subgroup $K$. By choosing a
  maximal abelian subspace $\afrak$ of $\pfrak$ the decomposition
  $\pfrak = \afrak \oplus \nfrak$ leads to factors $NAK$ of the group $G$,
 where $N$ is nilpotent
 and $A$ is abelian.
The map from $G$ to $X$ can be made explicit by choosing the maximal compact subgroup $K$ to
 be the isotropy group of a point $x_0$ in $X$, leading to $gx_0 = nak x_0 = nax_0$.

 In the remainder of this paper this framework will be
 illustrated in the context of modular inflation, with $j$-inflation as a particular example.

\section{Modular inflation}

The simplest theories of automorphic inflation can be formulated in the  context of
  classical modular functions and  forms, which are functions understood to be
  defined relative to arithmetic subgroups of  the modular group  $\rmSL(2,\mathZ)$,
 such as the Hecke congruence subgroups $\G_0(N)$ of level $N$, the principal congruence
 subgroups $\G(N)$, or other similar groups, collectively denoted by $\G_N$.
  The semisimple group $G$ in this case
  is $\rmSL(2,\mathR)$,
 and the maximal compact group $K=\rmSO(2,\mathR)$ leads to the domain $X$ which can
  be viewed as the upper halfplane $\cH = \rmSL(2,\mathR)/\rmSO(2,\mathR)$.

For modular forms $\Phi$ defined as group functions on $\rmSL(2,\mathR)$ the first
 automorphy constraint again identifies a character $\e_N$, possibly trivial, such that
 $\Phi(\g g) = \e_N(\g) \Phi(g)$.
 The second constraint, $K$-finiteness, can be made more precise in the modular case
 because the compact subgroup is abelian, hence its irreducible representations
 are one-dimensional. The resulting character determines the weight $w$ of the form via
 $\Phi(k_\theta g) = e^{iw\theta} \Phi(g)$, where $k_\theta$ is a rotation by $\theta$.
Finally, the differential constraint simplifies as well
 because the rank of $\rmSL(2,\mathR)$ is one, hence there is essentially only one
invariant differential operator, the Casimir element $\cC = h^{ij} X_iX_j$,
 where $h^{ij}$ is the inverse
 of the Killing metric $h_{ij}$, defined with respect to a basis $\{X_i\}$ of $\gfrak$ as
 $h_{ij} = \rmtr~\rmad_{X_i} \rmad_{X_j}$.
 The differential operator image $\Delta_\cC$ of $\cC$ is a
 multiple of the Laplace-Beltrami operator
 $\Delta_g = g^{-1/2} \del_i g^{ij} \sqrt{g} \del_j$.
 The eigenform constraint thus distinguishes between holomorphic and
 non-holomorphic forms.

 Modular forms $f(\tau)$ on the upper halfplane,
  characterized by their weight $w$, level $N$ and character $\e$,
  are obtained from group functions $\Phi$ on $\rmSL(2,\mathR)$ by considering the base point
 $x_0=i=\sqrt{-1}$ of the maximal compact subgroup $K$  as
  $f(\tau) = J(g,i) \Phi(g)$
 where $\tau=g\cdot i$.  For elements $\g$ in the discrete subgroup $\G_N$ of level $N$
 the automorphy factor is
 $J(\g, \tau) = (c\tau +d)^w$,
 where $\g$ is an element of $\G_N$ with rows $(a,b)$ and $(c,d)$,
 and the transformation behavior is given by
 \beq
  f(\g \tau) = \e_N(\g) (c\tau + d)^w f(\tau),~~~~~~ \g \in \G_N,
\eeq
 where $\rmdet(\g) =1$, and $\g\tau = (a\tau+b)/(c\tau+d)$
describes the fractional transform on the upper halfplane. The shift symmetry arises
 in the modular case
 from the generator given by $(a,b)=(1,1)$ and $(c,d)=(0,1)$, leading to the transformation
  that sends $\tau$
 to $\tau+1$. This shift symmetry also embeds into higher rank groups $G$ via the embedding of
 $\rmSL(2,\mathR)$ into $G$.

 The inflaton doublet  $\phi=(\phi^1,\phi^2)$ is parametrized by $\phi = \mu \tau$
 and the target space is equipped with the hyperbolic metric
 $ds^2 = d\tau d\otau/\left(\rmIm~\tau\right)^2$,
 i.e. the metric of the kinetic term is conformally flat with
 $G_{IJ} = \left(\mu/\phi^2\right)^2 \delta_{IJ}$.

Given a continuous group $G$, such as $\rmSL(2,\mathR)$, model building proceeds
 by first choosing a discrete subgroup $\G$, such as $\G_N \subset \rmSL(2,\mathZ)$.
This determines the level structure of the inflationary model.
 For a given  pair $(G, \G)$,  different modular functions can be obtained by considering quotients
  $f(\tau) = g(\tau)/h(\tau)$ of modular forms $g,h$ of the same weight with respect to
the same discrete group.
 Such
 functions then induce inflationary potentials of the type $V(f) = \Lambda^4 F(f,\of)$.
 This set-up of modular inflation thus  provides an extensive  framework in which
  finite-dimensional Hilbert spaces of modular forms
 can be  used to construct modular invariant potentials.
The example of $j$-inflation considered below
uses a particular modular function, constructed from
 modular forms with respect to the full modular group, but
other functions with respect to the full group can be considered, as well as forms
of higher level $N>1$.

\section{\bf $j$-Inflation}

The framework of modular inflation is exemplified in the following by considering an
 inflaton potential determined by a function invariant with respect to the
 full modular group. The space of classical modular forms with respect to
  $\rmSL(2,\mathZ)$ is in principle completely known since it is spanned by only two modular forms,
 the Eisenstein series $E_4(\tau)$ and $E_6(\tau)$ of weight 4 and 6 on the upper halfplane.
 These forms arise from the general Eisenstein series on the
 group $\rmSL(2,\mathR)$ defined as
 $E_s(g,f) = \sum_{\g}  f_s(\g g)$,
 where the sum is over a discrete quotient group, $s$ is a free parameter, $g$ is in
$\rmSL(2,\mathR)$, and $f_s(g)$ is a rescaled form of the function $f$ with a weight determined
 by $s$. The specific structure of $f_s$ encodes the form of the different
 Eisenstein series on the upper halfplane.
 The normalized Eisenstein series $E_w(\tau)$ of even weight $w$ that result from $E_s(g,f)$
 can be expressed in terms of the divisor function $\si_m(n) = \sum_{d|n} d^m$ as
 \beq
  E_w(\tau) = 1- \frac{2w}{B_w} \sum_n \si_{w-1}(n)e^{2\pi i n \tau},
\eeq
 where
  $\tau$ is in $\cH$ and $B_w$ are the Bernoulli numbers,
 which are related via Euler's formula to the Riemann zeta function as
 $2w!\zeta(w) = - (2\pi i)^w B_w$, with $w$ a positive even integer.
 The forms $E_4$ and $E_6$ are the unique elements of spaces of forms of weight
 four and six, respectively, up to normalization.

Holomorphic modular functions can be constructed from modular forms by considering quotients such that
 the denominator form is non-vanishing in the upper halfplane. A venerable modular form with this
property is the discriminant  function $\Delta$, the unique cusp form
 of weight twelve, up to normalization, defined by $\Delta(\tau) = \eta(\tau)^{24}$,
 where $\eta(\tau)$ is the Dedekind function
$\eta(\tau) = e^{2\pi i \tau/24} \prod_{n\geq 1} (1-e^{2\pi i n\tau})$.
 The form $\Delta$ arises in many different physical contexts,
 for example the partition function of the bosonic string, or as a geometric object,
 but is considered here simply as a building block of  modular
functions. Because $\Delta$ does not vanish on $\cH$,  modular functions without poles in $\cH$
 can be obtained by considering numerators of weight twelve. One of these
 is $E_4^3$, leading to the elliptic modular function
 \beq
  j(\tau) =\frac{ E_4(\tau)^3}{\Delta(\tau)},
 \eeq
  which is perhaps the most prominent modular function, going back to Kronecker's Jugendtraum,
 in particular his work on the class numbers of imaginary quadratic fields \cite{k1857},
   and Hermite's work on the quintic equation \cite{h1858}.
 Both $E_4$ and $\Delta$ are modular forms with respect to the full modular group,
 hence $j$ is a holomorphic function invariant under $\rmSL(2,\mathZ)$.

Given the function $j$  one can consider $j$-inflationary models defined by
 potentials $V=\Lambda^4 F(j,\oj)$.
 The simplest cases are obtained by setting e.g.  $F(f,\of) = (f\of)^p$ and in the following the focus
 will be on the model with $p=1$. As a first step toward a phenomenological analysis
 it is most convenient to consider the slow-roll approximation since this allows
 an analytic discussion of many aspects of the model, leaving a more precise numerical analysis
 of the exact dynamics to a more detailed discussion.
 Important parameters emphasized by the WMAP and {\sc Planck}
 analyses are given by the scalar spectral index $n_s$, defined in terms of the scalar
power spectrum $\cP_s(k)$ as
 $n_s-1 = d\ln \cP_s/d\ln k$,
 and the tensor-to-scalar ratio $r$, defined via the tensor power spectrum $\cP_t$ as
 $r = \cP_t/\cP_s$.
Similar to $n_s$ one considers the tensor power spectral index defined as
 $ n_t = d\ln \cP_t/d\ln k$.

 The power spectrum of the curvature perturbation
 can be written in multi-field inflation in terms of the number of
 $e$-foldings $N$ as \cite{ss95}
 \beq
 \cP_s ~=~ \left(\frac{H}{2\pi}\right)^2 G^{IJ}
 \frac{\del N}{\del \phi^I} \frac{\del N}{\del \phi^J},
 \eeq
while the tensor power spectrum was determined by Starobinsky \cite{s79} to be
 given by the Hubble parameter
 \beq
 \cP_t ~=~\frac{2}{\pi^2}  \left(\frac{H}{ M_\rmPl}\right)^2.
\eeq

For the inflationary model based on the $j$-function the parameters $n_s$ and $r$
 can be computed analytically in the slow roll approximation in
 terms of the dimensionless parameters $\e_I$ and $\eta_{IJ}$, defined as
 $\e_I = M_\rmPl V_{,I}/V$ and $\eta_{IJ} = M_\rmPl^2 \nabla_I \nabla_J V/V$,
 where $\del_I V = \del V/\del\phi^I$ and $\nabla_I$ denotes the covariant derivative in field
space. If the $\e_I$ are sufficiently small the universe accelerates.

In multi-field inflation  the spectral indices can be obtained in the slow-roll approximation as
 \beq
 n_s = 1- 3\e_I \e^I + 2 \frac{\eta_{IJ}\e^I\e^J}{\e_K\e^K}
 \eeq
 and
 $n_t = -\e_I \e^I$,
 while the tensor-to-scalar ratio  takes the form $r = -8n_t$.
 Here the indices are lowered and raised with the field space metric $G_{IJ}$ and its inverse.

 The observables $\cP_s(k_*)$, $n_s$ and $r$ can be expressed in the case of $j$-inflation in
 terms of the modular forms $E_4, E_6$ as well as the quasi-modular form $E_2$ by using
the relation
 \beq
 \e_I ~=~ -2\pi i^I \frac{M_\rmPl}{\mu} \left(\frac{E_6}{E_4} + (-1)^I \frac{{\bar E}_6}{{\bar E}_4}
  \right),
\eeq
where the index $I=1,2$ indicates the two components of $\tau = \tau^1+i\tau^2$.
 The height $\Lambda$ of the potential does not enter the scalar spectral index $n_s$ and
 the tensor ratio $r$,
 leading to expressions $n_s=n_s(\mu, E_w)$ and $r=r(\mu, E_w)$ with $w=2,4,6$.
 With these explicit results the spectral index and the tensor ratio can be evaluated in terms
 of the inflaton variable $\tau_* = \phi_*/\mu$ at horizon crossing
 in dependence of the scale $\mu$.
  Values of the dimensionless inflaton $\tau$ in the neighborhood of the zero $\tau=i$ of
  $E_6$ lead to $(\cP_s(k_*), n_s,r)$-parameters that are consistent with the observational
   results reported by the {\sc Planck} Collaboration \cite{planck13-inflation}.
 More precisely, the spectral index $n_s$ and a tensor-to-scalar ratio $r$ with values
 below the Planck
 bound can be obtained from a wide range of scales for $\mu$ around the Planck scale $M_\rmPl$.
 The scale $\Lambda$ that results from the scalar amplitude $\cP_s(k_*)$ includes the range
 $(10^{-5}-10^{-3})M_\rmPl$. For super-Planckian $\mu$ the number of e-foldings
 includes the standard interval between $N_e=50$ and $N_e=70$, with $n_s \cong 0.96*$,
 and the tensor ratio includes the range
 $0.02 \geq r \geq 10^{-4}$, which is within reach of proposed experiments,
  such as CMBPol \cite{cmbpol2} and PRISM \cite{prism13}.
  The values of the inflaton components $\phi^I_*$, determined by the choice
 of $\mu$, range from sub-Planckian to super-Planckian values for $\phi_*^1$,
 and generically super-Planckian values for $\phi_*^2$.

 It is similarly possible to constrain further parameters of these models for which
 WMAP and {\sc Planck} have provided limits. Among these are the running $\a_s$ of the
 scalar spectral index,  as well as  the non-Gaussianity parameter $f_\rmNL$
 of the bispectrum $B(\veck_1,\veck_2,\veck_3)$ and the parameters $\tau_\rmNL$ and
 $g_\rmNL$ of the trispectrum $T(\veck_1,\veck_2,\veck_3,\veck_4)$. These parameters
 can be expressed  in terms of the Eisenstein series as well,
 leading again to analytic expressions,
and the results can be compared with the limits imposed  by the
{\sc Planck} Collaboration \cite{planck13-24}. These results
involve higher derivative features of the potential and
 their more involved analysis, including the effects of the isocurvature perturbations,
 typically encoded in the transfer functions,
  is left to a more extensive future discussion.

The model of $j$-inflation is part of a class of functions comprised
 of genus zero modular functions. In the
 framework of modular inflation these provide a reservoir of different models that lend
themselves to a similar analysis. It is furthermore possible to
consider modular inflation based on more general
 forms of higher level by reducing the symmetry group to some congruence subgroup
 of level $N$.

\vskip .2truein

{\bf Acknowledgements.} \hfill \break
  It is a pleasure to thank Monika Lynker for discussions.  This work was supported in
  part by the NSF grant 0969875.
  The author is grateful for the hospitality and support by the
  Simons Center for Geometry and Physics, where part of this work was done.

\end{document}